\documentclass[journal]{IEEEtran}
\usepackage{algorithm}
\usepackage{algpseudocode}
\usepackage{hyperref}

\newcommand{\myref}[1]{\href{#1}{#1}}

\ifCLASSINFOpdf
  \usepackage[pdftex]{graphicx}
\else
  \usepackage[dvips]{graphicx}
\fi
\usepackage{booktabs}
\usepackage{color,soul}

\usepackage{amsmath}

\usepackage{enumitem}

\usepackage{glossaries}
\setacronymstyle{long-short}
\glsdisablehyper
\newacronym{5gaa}{5GAA}{5G Automotive Association}
\newacronym{adas}{ADAS}{Advanced Driver-assistance System}
\newacronym{atcll}{ATCLL}{Aveiro Tech City Living Lab}
\newacronym{bl}{BL}{baseline}
\newacronym{cam}{CAM}{Cooperative Awareness Message}
\newacronym{cfs}{CFS}{Completely Fair Scheduling}
\newacronym{cv2x}{C-V2X}{Cellular-V2X}
\newacronym{cgroup}{Cgroup}{Control group}
\newacronym{c-its}{C-ITS}{Cooperative Intelligent Transport Systems}
\newacronym{ctlta}{CTLTA}{Cross-Traffic Left-Turn Assist}
\newacronym{ccdf}{CCDF}{Complementary Cumulative Distribution Function}
\newacronym{ebw}{EBW}{Emergency Brake Warning}
\newacronym{edf}{EDF}{Earliest Deadline First}
\newacronym{gn}{GN}{GeoNetworking}
\newacronym{hdss}{HDSS}{Automated Driving Lane Change}
\newacronym{hf}{HF}{High-Frequency}
\newacronym{hp}{HP}{High Priority}
\newacronym{ims}{IMS}{Intersection Movement Assist}
\newacronym{iot}{IoT}{Internet-of-things}
\newacronym{ipc}{IPC}{Inter-process communication}
\newacronym{its}{ITS}{Intelligent Transportation Systems}
\newacronym{kasan}{KASAN}{Kernel Address Sanitizer}
\newacronym{stibp}{STIBP}{Single Thread Indirect Branch Predictors}
\newacronym{lf}{LF}{Low-Frequency}
\newacronym{lp}{LP}{Low Priority}
\newacronym{mapem}{MAPEM}{MAP Extended Message}
\newacronym{vam}{VAM}{VRU Awareness Message}
\newacronym{mec}{MEC}{Multi-Access Edge Computing}
\newacronym{nd}{ND}{Native as Docker}
\newacronym{obu}{OBU}{On-Board Unit}
\newacronym{os}{OS}{Operating System}
\newacronym{rsu}{RSU}{Road Side Unit}
\newacronym{rt}{RT}{Real-Time}
\newacronym{rtos}{RTOS}{Real-Time Operating System}
\newacronym{rtsa}{RTSA}{Real-Time Situational Awareness}
\newacronym{sbc}{SBC}{Single-Board Computer}
\newacronym{sdk}{SDK}{Software Development Kit}
\newacronym{slr}{SLR}{Service Level Requirements}
\newacronym{soc}{SoC}{System on a Chip}
\newacronym{sp}{SP}{Same Priority}
\newacronym{stp}{STP}{See-Through for Passing}
\newacronym{tc}{TC}{Traffic Control}
\newacronym{urllc}{URLLC}{Ultra-Reliable and Low-Latency}
\newacronym{v2x}{V2X}{Vehicle-to-everything}
\newacronym{vanet}{VANET}{Vehicular Ad-hoc Network}
\newacronym{vec}{VEC}{Vehicle Edge Computing}
\newacronym{vhp}{VHP}{Very High Priority}
\newacronym{vm}{VM}{Virtual Machine}
\newacronym{vru}{VRU}{Vulnerable Road User}
\newacronym{wcet}{WCET}{Worst-Case Execution Time}
\newacronym{cpu}{CPU}{Central Processing Unit}
\newacronym{ram}{RAM}{Random Access Memory}
\newacronym{gpu}{GPU}{Graphics Processing Unit}
\newacronym{rtt}{RTT}{Round-Trip Time}
\newacronym{dnat}{DNAT}{Destination Network Address Translation}
\newacronym{grpc}{gRPC}{Google Remote Procedure Call}
\newacronym{api}{API}{Application Programming Interface}
\newacronym{dns}{DNS}{Domain Name System}
\newacronym{ip}{IP}{Internet Protocol}
\newacronym{fifo}{FIFO}{First-In First-Out}
\newacronym{cri}{CRI}{Container Runtime Interface}
\newacronym{cni}{CNI}{Container Network Interface}
\newacronym{cli}{CLI}{Command Line Interface}
\newacronym{http}{HTTP}{Hypertext Transfer Protocol}
\newacronym{csv}{CSV}{Comma-Separated Values}
\newacronym{tcp}{TCP}{Transmission Control Protocol}
\newacronym{udp}{UDP}{User Datagram Protocol}
\newacronym{url}{URL}{Uniform Resource Locator}
\newacronym{lan}{LAN}{Local Area Network}
\newacronym{mqtt}{MQTT}{Message Queuing Telemetry Transport}
\newacronym{dds}{DDS}{Data Distribution Service}
\newacronym{etsi}{ETSI}{European Telecommunications Standards Institute}
\newacronym{json}{JSON}{JavaScript Object Notation}
\newacronym{yaml}{YAML}{Yet Another Markup Language}
\newacronym{QoS}{QoS}{Quality of Service}
\newacronym{gps}{GPS}{Global Positioning System}
\newacronym{posix}{POSIX}{Portable Operating System Interface}
\newacronym{sysv}{SysV}{Unix System V}
\newacronym{rssi}{RSSI}{Received Signal Strength Indication}
\newacronym{cuda}{CUDA}{Compute Unified Device Architecture}
\newacronym{cr}{CR}{Custom Resource}
\newacronym{crd}{CRD}{Custom Resource Definition}
\newacronym{vcs}{VCS}{Version Control System}

\begin{document}
%
\title{Edge-Cloud Continuum Orchestration of Critical Services: A Smart-City Approach}
%
%
%
\author{\IEEEauthorblockN{
Rodrigo Rosmaninho\IEEEauthorrefmark{1}\IEEEauthorrefmark{2},
Duarte Raposo\IEEEauthorrefmark{1}, 
Pedro Rito\IEEEauthorrefmark{1}, 
Susana Sargento\IEEEauthorrefmark{1}\IEEEauthorrefmark{2}\\}
\IEEEauthorblockA{\IEEEauthorrefmark{1}Instituto de Telecomunica\c{c}\~{o}es, 3810-193 Aveiro, Portugal\\}
\IEEEauthorblockA{\IEEEauthorrefmark{2}DETI, University of Aveiro, 3810-193 Aveiro, Portugal}
}

%



\maketitle

\begin{abstract}
Smart-city services are typically developed as closed systems within each city's vertical, communicating and interacting with cloud services while remaining isolated within each provider's domain. With the emergence of 5G private domains and the introduction of new M2M services focusing on autonomous systems, there is a shift from the cloud-based approach to a distributed edge computing paradigm, in a \textit{continuum} orchestration. However, an essential component is missing. Current orchestration tools, designed for cloud-based deployments, lack robust workload isolation, fail to meet timing constraints, and are not tailored to the resource-constrained nature of edge devices. Therefore, new orchestration methods are needed to support MEC environments. The work presented in this paper addresses this gap. Based on the real needs of a smart-city testbed - the Aveiro Living Lab-, we developed a set of orchestration components to facilitate the seamless orchestration of both cloud and edge-based services, encompassing both critical and non-critical services. This work extends the current Kubernetes orchestration platform to include a novel location-specific resource definition, a custom scheduler to accommodate real-time and legacy services, continuous service monitoring to detect sub-optimal states, and a refined load balancing mechanism that prioritizes the fastest response times. 
\end{abstract}


%
\IEEEpeerreviewmaketitle

\section{Introduction}

There is a growing need for comprehensive orchestration mechanisms to efficiently manage the growing scale and complexity of future smart-city edge computing infrastructures. Historically, smart-city services were developed as closed systems specific to each city's verticals \cite{kirimtat_2020}, wherein sensors, networks, and processing devices remained isolated within each service provider's domain. This approach incurred additional capital and operational expenditures, as it involved duplicating infrastructure across providers, financed by public funds. The emergence of 5G private networks \cite{Sulieman2022}, coupled with advancements in vehicular networks featuring autonomous vehicles \cite{Thandavarayan2020} and smart mobility solutions, has catalyzed a shift towards centralizing cloud services back to a distributed edge computing paradigm. This shift has the potential to reshape the current landscape, allowing public infrastructure to be shared among multiple providers and thereby reducing costs.

Nonetheless, a pivotal element essential to materialize this concept involves the orchestration of smart-city edge services, similar to the management of cloud services. This endeavor demands a significant overhaul of current orchestration systems (primarily fine-tuned for cloud deployments) to confront the distinctive challenges posed by edge services. These solutions also lack robust workload isolation mechanisms, which are important in this context since smart-cities support a wide range of use cases, some of which are of a critical nature and need to comply with timing constraints, despite the potential interference from other co-located services. This is further exacerbated by the resource-constrained nature of the computing nodes. If a practical solution to these issues could be found, smart-city deployments could transition from multiple, isolated, provider-owned infrastructures (eg: ISPs, Utilities, Environmental Monitoring, Public Transit, etc) to a single cohesive Multi-access Edge Computing (MEC) environment capable of ensuring the fulfillment of critical time constraints while still providing computing headroom to other stakeholders and use cases. This would increase efficiency and control and enable a wider range of collaboration and data fusion opportunities, while potentially reducing overall expenditure.

Based on the needs of the Aveiro Living Lab \cite{atcll}, and in the lack of tools to support such environments, this paper introduces innovative orchestration techniques that could be used to manage the specificities of smart-city edge services. Our proposal includes: 1) a novel resource definition enabling the creation of location-specific services; 2) a custom scheduler with four integrated plugins to facilitate the orchestration of real-time services and enhance service dependency management; 3) a set of runtime extensions designed to manage legacy applications and to assign real-time CPU scheduling policies to services; 4) continuous service monitoring of the cluster state to detect sub-optimal conditions and the need for re-scheduling; 5) a refined load balancing method that transitions from conventional cloud balancing, which aims for maximum fairness and balanced resource usage between replicas, to an approach that prioritizes the fastest response time. Our results demonstrate that the new techniques significantly enhance current cloud platforms, such as Kubernetes, in supporting edge-based orchestration across multiple domains.

\begin{figure}[ht]
	\centering
	\includegraphics[width=\columnwidth]{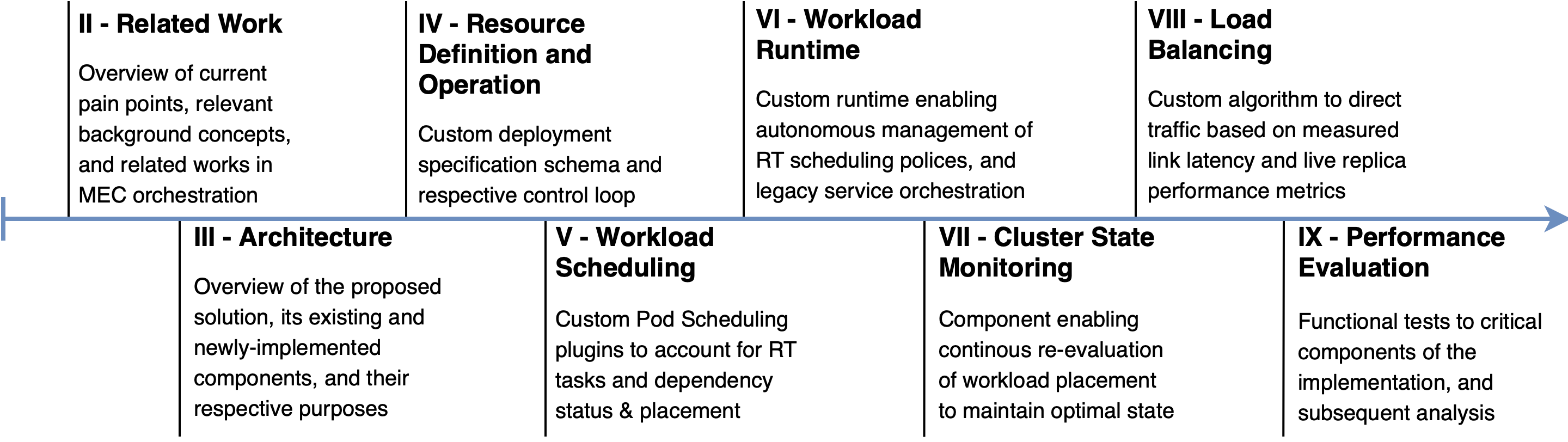}
	\caption{High-level overview of the structure of this work}
	\label{fig:paper_arch}
    \vspace{-5mm}
\end{figure}

The remaining of this article is organized as follows (see figure \ref{fig:paper_arch}). Section \ref{related} provides an overview of the related work, by exploring orchestration solutions that addresses the several domains explored in our work. Section \ref{sec:architecture} introduces the edge-based architecture, and its new components. Section \ref{sec:crd} describes the new resource definition and its operation. Section \ref{sec:scheduler} presents the new workload scheduling, which integrates real-time scheduling and the dependency orchestration. Section \ref{sec:runtime} covers workload runtime, presenting the extensions the support of legacy applications and CPU scheduling policies. Section \ref{sec:cluster_state_monitor} the edge-cluster state monitoring. Section \ref{sec:load_balancer} outlines the new load-balancing mechanism. Lastly, section \ref{sec:performance_evaluation} assesses the performance evaluation of the new components. The article concludes with section~\ref{sec:conclusion}, summarizing the key findings and new directions for future work.

\section{Related Work}
\label{related}
\gls{mec} topologies bring communication, storage, and computational capabilities closer to the end-users, potentiating a reduction of network load and latency, and consequently enabling real-time operations. Edge devices can collaborate to perform functions like data collection, processing, model training, caching, and data analytics. This is achieved through methods such as edge caching, edge training, and edge offloading, which enhance the capabilities and performance of \gls{iot} services at the network edge \cite{Zhang2022}.

Orchestration solutions (like, for instance, Kubernetes) typically involve a cluster of compute nodes managed by a centralised orchestrator with the capabilities to deploy, monitor, migrate and scale services, and load-balance requests. However, since Kubernetes was developed primarily for cloud computing environments, it is not fully suited for use in resource-constrained \gls{mec} topologies. Some of the challenges associated with bringing edge computing to production environments are: \gls{QoS}; data management; network scalability; security \& privacy; integration in the future 5G and beyond ecosystem; and application placement \& resource allocation \cite{Mendiboure19}. 
A sizeable portion of ongoing research is focused on this last consideration, since the computational capabilities of \gls{mec} topologies are often limited due to resource-constrained hardware coupled with high demand for workload placement. Moreover, some applications benefit from mixed deployments in both cloud and edge computing environments. Further research is needed regarding efficient orchestration strategies to handle this heterogeneous scenario \cite{Darwish18}. 
For instance, in \cite{Loghin19} the authors concluded that there is no ideal approach, and the decision between edge-only, hybrid, or cloud-only should be made according to the characteristics of each individual application. In \cite{Zhang22} authors study this issue, taking into account the maximum acceptable delay. In \cite{Lai18}, the authors propose a framework based on fog nodes to efficiently monitor the network, avoiding extraneous processing. \cite{9586998, 9277773, 9348695, 10019879} propose different algorithms and techniques to optimise \gls{QoS}, offloading efficiency, and cluster resource allocation, based partly on the emerging Fog Computing paradigm.
In general, these approaches are focused solely on optimising initial workload allocations, and don't consider the possibility of continuous re-optimisation, depending on the evolution of cluster conditions over time.

 In \cite{Santos2019}, the authors analyse the various scheduling features available in Kubernetes and present an extension to the default Kubernetes scheduler in order to make the scheduling process aware of network bandwidth and latencies. Results show that the proposed network-aware scheduler can significantly improve the service provisioning of the default scheduler by achieving a reduction of 80\% in terms of network latency. \cite{phd_loadbalancer} proposes a similar solution, with the notable addition of a customised load-balancer. However, such an approach should ideally take the processing latency introduced by the service into consideration as well. In \cite{Javed2020} the authors use Kubernetes as a fault-tolerant framework to run \gls{vanet} applications, in scenarios where one or more RSU worker nodes (or their communication links) could unexpectedly fail. The authors only focus their proposal on the impact of fault-tolerant mechanisms in terms of latency and throughput. Lastly, from a more architectural perspective, the work presented in \cite{Bohm2022} discusses the usage of Kubernetes as an orchestration platform for Edge Computing and presents some limitations with regard to resource awareness and architectural shortcomings, as well as potential solutions for each one. These solutions include the use of lightweight Kubernetes distributions, custom metric servers, and Scheduler extensions. 

When attempting to guarantee stable and predictable task execution times, we must also consider how the underlying operating system schedules the allocation of CPU time between all the active threads.
Since Linux is a general purpose operating system, by default, tasks are scheduled using the \gls{cfs} algorithm \cite{Struhar2020}, which strives to maintain fairness. However, this approach can make critical services vulnerable to resource starvation \cite{Madej2020}, causing unexpected delays. 
Fortunately, Linux also provides native support for a set of \gls{rt} scheduling policies \cite{Struhar2020}, which enable the prioritisation of critical workloads and even the preemption of CPU cycles from lower priority tasks, greatly improving the stability of the former's performance. \cite{artigo_NOMS} analyses the impact of these and other mechanisms on the performance of a critical \gls{v2x} application deployed on resource-constrained hardware used on a real \gls{mec} cluster \cite{atcll}, and shows that they are very effective even when coupled with container-based deployments. 
The survey in \cite{Struhar2020} details the current research efforts regarding the application of these \gls{rt} policies in containerised environments, and the relevant management mechanisms. However, these efforts are mainly focused on industrial automation use cases instead of more generic \gls{mec} scenarios, and tackle the issue at the level of an individual node rather than of a centralised orchestrator. This is the case with \cite{8416214, 8812232, abeni19}, which propose a hierarchical scheduling approach at the Linux kernel level. 
\cite{9613685} is the most similar to the goals of our work, since it presents proof-of-concept adaptations to Kubernetes based on a scheduler plugin and node-level \gls{rt} management daemon. However, it does not include certain features such as the preemption of non-\gls{rt} and lower-priority Pods at the Kubernetes scheduler level. 

In a broader sense, we found that there is a lack of research taking a holistic approach to the overall problem by addressing these issues simultaneously in an integrated architecture, like our work proposes.




\section{Architecture}
\label{sec:architecture}

This section details the architecture of the proposed orchestration framework that provides continuum dynamic orchestration of edge resources for services according to their requirements and location. Firstly, the background and the requirements of the new components are explored. Thus, each module is explained in a generic manner, without restricting to any specific existing implementation (i.e., Kubernetes, Docker Swarm, etc). In fact, the solutions presented in this architecture are applicable to most orchestration frameworks.
From a functionality standpoint, the architecture can be divided into 3 distinct groups, which are described in the following subsections.

\subsection{Background \& Requirements}
\label{sec:background}

To be able to integrate time-sensitive services in this new edge platform, new features need to be mapped into the architecture. In specific, (1) automatic configuration of real-time scheduling parameters for critical processes; (2) workloads to nodes based on customised algorithms that take into account latency and application-level metrics of their dependencies; (3) migration of workload allocations that become sub-optimal over time; and (4) load balancing mechanisms based on customised algorithms, that take into account the latency and application-level metrics of the destination replicas.
The objective of this work is not to implement an entirely new orchestrator, since that would prove a time-consuming and ultimately ineffective process. Instead, the proposed functionalities and integrations should, whenever possible, be implemented as extensions of the various relevant components of existing orchestration solutions, either by following the plugin design pattern or by making extensive use of available control plane APIs and SDKs. 

The process of selecting a worker node for a given deployment should take into account the real-time scheduling attributes of both the candidate application and the services that are already in execution, if any. That way, \gls{rt} services can be scheduled in a balanced manner that minimizes their interference on regular services.  
The scheduler should also compute the feasibility of assigning the deployment to a given node, in terms of quota utilisation, so as to prevent the over-allocation of \gls{rt} workloads relative to the actual capacity of the node.
In scenarios where a deployment specifies one or more dependencies, the scheduling process should give preference to nodes that minimize the network latency between those services. This dependency-aware algorithm should also take into account the current performance level of each of the replicas of a dependency, thus preferring the most-performing replicas and maximising the quality of service. 


In most orchestration solutions, if Kubernetes is used, Pods are scheduled to the most optimal worker node at the given moment, and this allocation is not re-evaluated after the initial scheduling process. This means that the cluster will inevitably deteriorate into a sub-optimal state.
To address this, the system should continuously monitor active deployments and proactively migrate them to other nodes that increase their performance and/or the overall cluster balance. This logic should take into account all of the complex factors included in the original scheduling approach, to avoid spurious migrations.


In order to take advantage of the dependency-aware scheduling algorithm, the system shall include an intelligent load-balancing algorithm that gives precedence to destinations that incur the lowest amount of latency when forwarding network requests to replicas.
Additionally, the current performance levels of the replicas shall also be taken into account, so as to increase the overall stability and quality of service.

\begin{figure}[b!]
    \centering
    \includegraphics[width=\columnwidth]{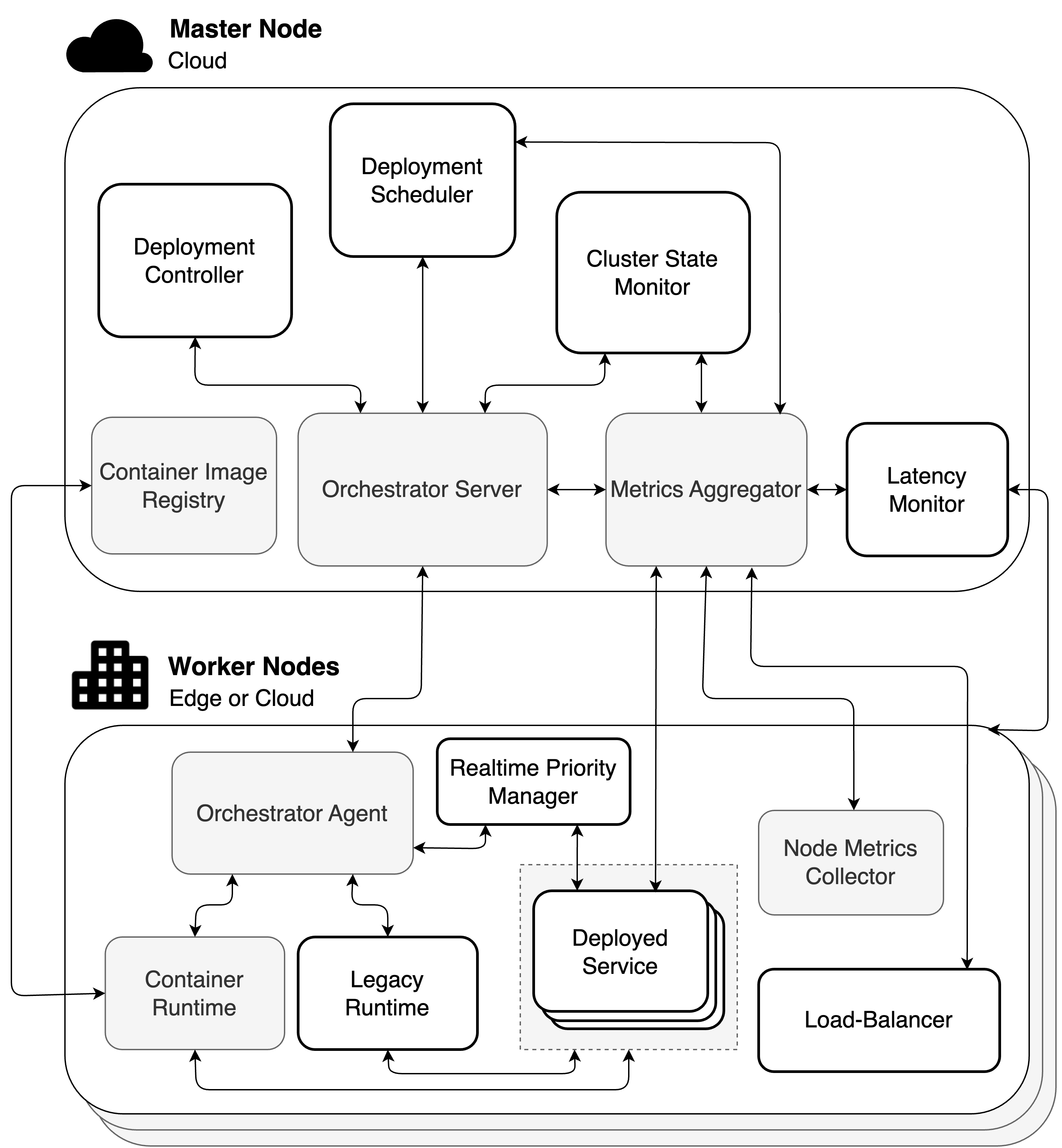}
    \caption{Overview of the proposed architecture}
    \label{fig:proposed_architecture}
\end{figure}

\begin{figure}[b!]
    \centering
    \includegraphics[width=\columnwidth]{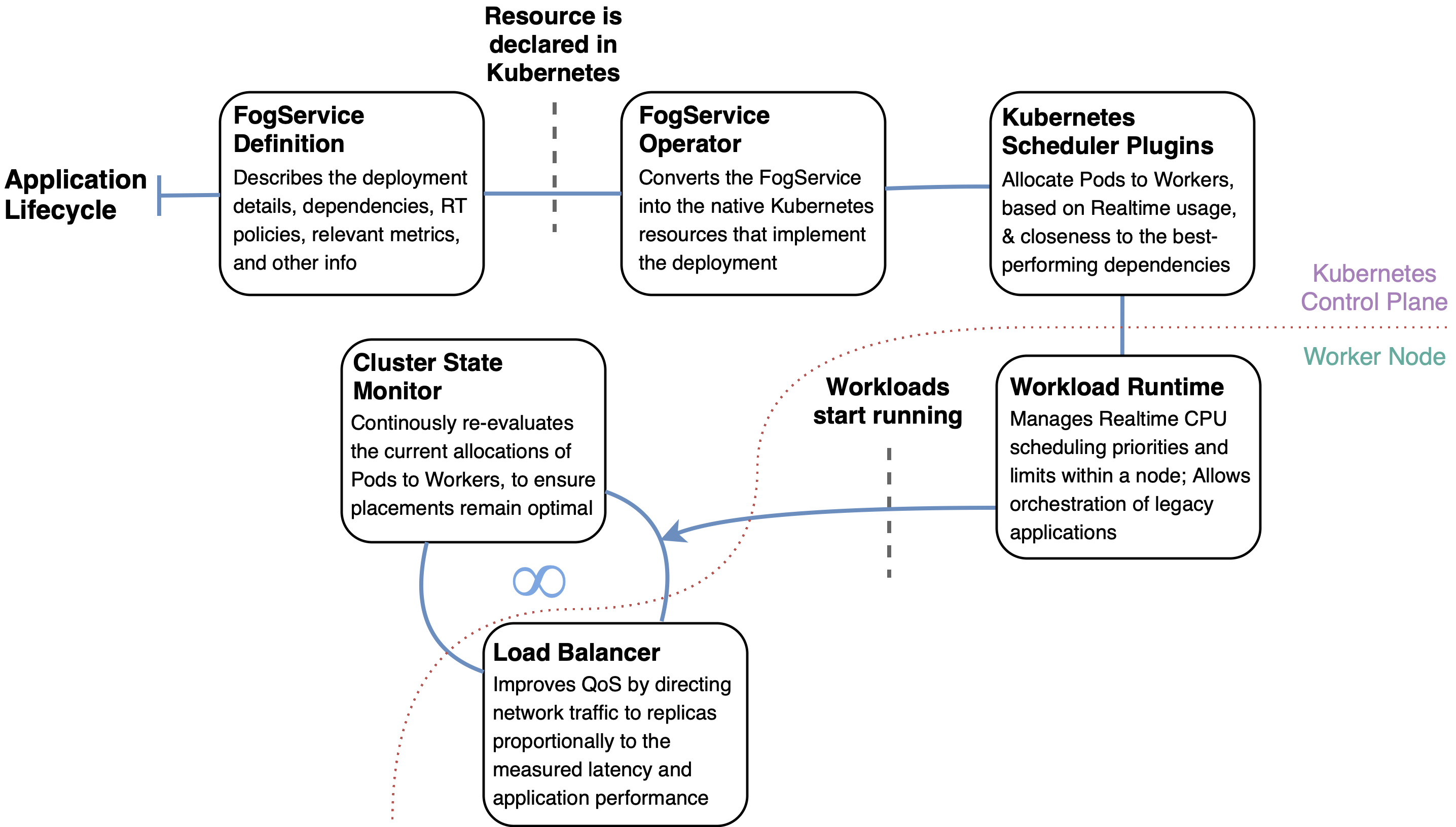}
    \caption{Overview of the lifecycle of a deployment, and the role of the critical components of this implementation}
    \label{fig:lifecycle}
\end{figure}

Given its extensional nature, the proposal includes a mix between components that introduce new features envisioned for the purpose of this work, and industry-standard components that already have several implementations available for use. 
In order to distinguish between the two types of modules more clearly, the former are pictured as highlighted blocks in figure \ref{fig:proposed_architecture}.  
Figure \ref{fig:lifecycle} presents the high-level lifecycle of a deployment using this architecture, and how the various components of the solution interact, optimize, and implement the deployment.


\subsection{Orchestration}

This includes all the components that contribute to the service deployment pipeline, from the moment that it is requested, to when the service is successfully executing in a worker node with the desired real-time priority applied, if applicable:

\begin{itemize}[leftmargin=*]
	\item \textbf{Orchestrator Server} - Manages and monitors all deployments and worker nodes on the cluster. It exposes an \gls{api} that services can use to request a deployment and consult its state. 
	\item \textbf{Deployment Controller \textit{(new)}} - Interprets the customised Deployment Specification File and converts it into a set of actions that the orchestrator server must perform in order to successfully complete the deployment.
	\item \textbf{Deployment Scheduler \textit{(new)}} - Each time a new deployment is queued, this component analyses its requirements and the current state of the cluster in order to select the most appropriate worker node to allocate it to. This selection is done based on several factors, including the current value of crucial node metrics, the \gls{rt} feasibility of the node's task set, and the node's proximity to the deployment's dependencies. 
	\item \textbf{Orchestrator Agent} - Interfaces with the Orchestrator Server and with the available runtimes in order to create, manage, and monitor local deployments. 
	\item \textbf{Container Runtime} - Manages the complete lifecycle of a containerised deployment, including downloading the respective image, creating and executing the container, and exposing its logs and current status to the orchestrator.
	\item \textbf{Container Image Registry} - Provides centralised storage of container images for every deployment. Images are uploaded by developers and administrators, and downloaded by the Worker nodes after a new deployment is scheduled.
	\item \textbf{Legacy Runtime \textit{(new)}} - Manages the complete lifecycle of a legacy deployment, including downloading and installing dependencies, executing the relevant processes, and exposing their logs and current status to the orchestrator.
	\item \textbf{Realtime Priority Manager \textit{(new)}} - Interfaces with the worker node's operating system in order to apply the desired realtime priority or SCHED\_DEADLINE attributes to the processes of a newly assigned deployment. It is also capable of dynamically re-assigning the priorities of existing deployments, if necessary.
\end{itemize}

\subsection{Metrics Aggregation}

This includes the components that comprise the metric collection, persistence, and querying architecture of the proposed system:

\begin{itemize}[leftmargin=*]
	\item \textbf{Metrics Aggregator} - Continuously pulls the most recent values from the various metric sources and persists them in a database, thus allowing the remaining architectural components to perform complex queries for both current values and historical trends. 
	\item \textbf{Deployed Service \textit{(new)}} - When applicable, it exposes application-level metrics that provide insight into the performance of the application and allow the orchestrator to perform corrective actions when it enters a degraded state.
	\item \textbf{Node Metrics Collector} - Exposes a vast array of detailed metrics regarding the state of the worker node and the utilisation of its resources (\gls{cpu}, \gls{ram}, Storage, Networking, among others).
	\item \textbf{Latency Monitor \textit{(new)}} - Continuously collects and exposes the current ping averages for every node in the cluster, thereby enabling other components to make informed decisions regarding the state of the network.
\end{itemize}

\subsection{Monitoring and Load-Balancing}
 
This includes other components that do not fit in the previous groups, but play a critical role in assuring quality of service and stability over time:
 
\begin{itemize}[leftmargin=*]
	\item \textbf{Cluster State Monitor \textit{(new)}} - Continuously monitors the current cluster state in order to correct deployment allocations that have gradually become sub-optimal by evicting the respective deployments, thereby forcing their re-scheduling into the most suitable node. 
	\item \textbf{Load-Balancer \textit{(new)}} - Balances network requests to a service between all the running replicas of that service. When possible, this process takes into account the current network link latency between the client and the potential destination, and additionally, the application-level metrics exposed by the replicas in order to give precedence to the replicas which will provide the best Quality of Service.
\end{itemize}

\section{Resource Definition and Operation}
\label{sec:crd}



Some of the additions proposed in this work require configuration parameters that are not supported by the current Kubernetes \gls{api}, such as Realtime CPU Scheduling attributes to apply, or application-level metrics to monitor. To address this, the proposed architecture includes a new type of service descriptor file named FogService, that abstracts the usual Kubernetes complexity behind a single file that supports both the most commonly used Kubernetes configurations, and all the additions required by our proposed features.



When a new FogService resource is declared, the respective FogService controller parses each of its sections and autonomously creates and declares all the different built-in Kubernetes resources that the service requires, thus allowing Kubernetes to proceed with the orchestration of the deployment, and fulfilling FogService's role as an abstraction layer for the default resources.

In cloud-based cluster environments, the existence of multiple worker nodes is primarily used to provide horizontal scaling (and, thus, high-availability, load-balancing, and increased performance) to generic applications. In some specific cases, the worker nodes can also be geographically dispersed, in order to increase the proximity of end clients to an available replica of a given service.
Therefore, the replication controllers used by Kubernetes' native resources are optimised for the requirements of these specific use cases.

As previously discussed, edge-computing clusters are also capable of fulfilling these use cases, but normally have the advantage of providing even greater proximity between an application and its clients.
However, in the context of a smart-city, each node typically has the possibility of being equipped with a set of sensors and communication technologies that allow it to collect and process data relevant to its specific location, and interface with the real world.   
As such, these types of clusters can also implement a slightly different usage pattern where a given service is deployed to several different nodes, but each one is conceptually treated as a distinct stateful application whose scope of operation is limited to that specific location.
Examples of this type of usage pattern include, for instance:
\begin{itemize}[leftmargin=*]
	\item An object detection service using video frames obtained in real-time from an \gls{ip}-enabled camera;
	\item A local database that records the data points generated at a given location, so that other services can access them;
	\item A vehicular network Road-Side Unit service that sends and receives C-ITS messages over ITS-G5.
\end{itemize}
These examples constitute services that are meant to be deployed to multiple distinct locations/nodes, but do not fit the archetypal pattern of being replicas of a single cluster-scoped application. Instead, their deployment is implemented as a set of location-scoped applications that utilise location-specific resources and produce/consume location-specific data. 
Following this approach, each location-scoped service can then have its own custom configuration, which is important since a given application may require different configuration options depending on the location it is allocated to, as each worker node can typically have a distinct set of sensors and/or communication interfaces that best suit its specific location and use cases.
Furthermore, a given location-scoped service can also have its own set of replicas which keep the original's scope. This enables the operator to perform differential scaling of replicas on each location-scoped deployment, allowing it to react to higher utilisation levels or performance issues in a specific location.
Notably, these replica Pods can potentially be scheduled on different nodes due to scheduling constraints, while still keeping their original location scope (provided that the application does not require access to specialised resources).

\section{Workload Scheduling}
\label{sec:scheduler}

The proposed custom scheduler is implemented as a set of plugins which augment the default scheduling framework with new capabilities, while keeping all the existing tried-and-true features that developers expect. These plugins are registered as Filter, PostFilter, and/or Score extension points within the Pod scheduling lifecycle.

\subsection{Realtime}
\label{sec:realtime_plugin}

Given that the default Kubernetes scheduler was not designed to take a Pod's Realtime attributes into account, Realtime Pods are scheduled using the same criteria as any other Pod. In terms of resource balancing, this means that Pods are preferably allocated to the node with the smallest percentage of resources claimed by existing Pods, and no effort is made to minimise the \gls{cpu} utilisation of real-time processes. As a result, some nodes will inevitably exhibit a higher number of Realtime Pods than others, which is sub-optimal since RT-enabled processes can have a significant impact on the performance of regular processes even when there is \gls{cpu} headroom available, by virtue of their precedence in terms of \gls{cpu} time allocation, and the preemption mechanism that interrupts other tasks. 

 To address these issues, the \textit{Realtime} scheduler plugin implements the Score extension point in order to rank candidate nodes by their overall share of \gls{cpu} capacity being allocated to Realtime tasks.
 Since Pods can spawn a mix of both Realtime and regular processes, the \gls{rt} utilisation of a Pod is not necessarily equal to its total \gls{cpu} resource requests, and must instead be calculated from other fields included in the FogService specification:
 
 
 
 \[ U = \sum_{i=1}^{n} \frac{runtime_i}{period_i} + \sum_{i=1}^{k} requests_i  \] where \textit{U} represents the overall utilisation, \textit{n} represents the number of SCHED\_DEADLINE processes running on the worker node, and \textit{k} represents the number of SCHED\_FIFO processes running on the worker node.

 
 This final utilisation value can then be used to allocate a score to the node, thus fulfilling the plugin's main objective.
However, while this approach successfully minimizes the overall Realtime interference on each node, it does not address the interference that higher-priority \gls{rt} tasks have on their lower-priority counterparts.
In fact, any SCHED\_DEADLINE task can potentially become a source of interference for any other tasks, depending on runtime circumstances that are not trivial to predict at the Kubernetes Scheduler level, especially for multi-core systems.

It is important to note that the available capacity for regular processes and \gls{rt}-enabled processes are not necessarily identical. In fact, Linux systems have the option to set a limit on the quota of the overall processing time that can be allocated to Realtime tasks, which is typically set at 95\% by default so that \gls{rt}-enabled processes cannot completely starve out critical Kernel and Operating System functions.
Therefore, Kubernetes' default approach is not entirely suitable in this context, since it could result in the allocation of a Realtime Pod to a node with sufficient overall capacity but insufficient Realtime capacity. 

To ensure that worker nodes do not become over-allocated, the Realtime plugin implements the Filter extension point in order to perform a utilisation analysis of each node and preemptively exclude any nodes where the prospective Realtime Pod would exceed the remaining capacity for \gls{rt}-enabled tasks. The aforementioned Realtime utilisation quota is set on a per-node basis using the \textit{sched\_rt\_period\_us} and \textit{sched\_rt\_runtime\_us} kernel parameters and, as such, may not be identical for every node, especially in heterogeneous clusters with distinct hardware constraints. 
For this reason, the approach presented in this work requires each worker node to include the values of its \textit{sched\_rt\_period\_us} and \textit{sched\_rt\_runtime\_us} kernel parameters as labels within the node resource.


The resulting algorithm can be presented in mathematical terms using the following formula:



\begin{multline}
    \sum_{i=1}^{n} \frac{runtime_i}{period_i} + \sum_{i=1}^{k} requests_i + \sum_{i=1}^{j} \frac{runtime_i}{period_i} + \\
    \sum_{i=1}^{w} requests_i \le M *  \frac{sched\_rt\_runtime\_us}{sched\_rt\_period\_us}
\end{multline}

where \textit{n} represents the number of SCHED\_DEADLINE processes running on the node, \textit{k} represents the number of SCHED\_FIFO processes running on the worker node, \textit{j} represents the number of SCHED\_DEADLINE processes spawned by the prospective Pod, \textit{w} represents the number of SCHED\_FIFO processes spawned by the prospective Pod, and \textit{M} represents the number of processor cores installed on the host.
In extreme scenarios where there are no available nodes with sufficient resources for a given candidate Pod, the plugin will remove every node from consideration, thus marking the Pod as \textit{Unschedulable} even if there are lower-priority \gls{rt} Pods currently running. 
To avoid this, the plugin also implements the PostFilter extension point in order to perform a preemption analysis of all the worker nodes and evict the lowest priority Pod(s) that will free sufficient resources for the allocation of the higher-priority candidate Pod.

\subsection{Dependencies}

Kubernetes includes native support to establish a dependency relation between two services by including a PodAffinity section within the deployment resource. When this relation exists, the scheduler automatically favours nodes that possess a replica of the service on which the prospective Pod depends. In cases where this is not possible, it attempts to allocate the Pod to a node belonging to the same topology zone as nodes where dependencies are located, thus leveraging the fact that zones frequently represent sets of nodes in geographical proximity, which should therefore be able to communicate with lower latency. Nevertheless, this approach is not able to take into account more complex network topologies, or situations where it is not possible to allocate a Pod to any node within the same topology zone as a dependency. In these scenarios, the scheduler's ranking of worker nodes can become, in effect, arbitrary, when evaluated from the standpoint of achieving closeness to dependencies. 
Furthermore, in situations where there are multiple replicas of the same dependency running on separate nodes, the scheduler should also be able to account for differences in processing performance. Ideally, Pods should be scheduled as close as possible to the least-used and/or best-performing replicas of a dependency (as determined by one or more predefined metrics), instead of any arbitrary replica.

Before the final node score can be calculated, the scores of each dependency's replicas are aggregated into a single value that represents the overall quality of the communication experience between the prospective Pod and the dependency, on the node being considered.
This calculation takes into account the customised load-balancer implementation included in this work, that ensures that the highest ranked replicas receive the majority of requests.
As will be discussed later in section \ref{sec:load_balancer}, the behaviour of this load-balancing algorithm is based on a Markov chain where each state represents a candidate replica, and the state transition probabilities reflect the differences in metric values and link latency. Each time a new network request is received, the load balancer uses the chain to select a suitable destination replica. 
Therefore, in order to provide the best possible approximation of real-world behaviour, the \textit{Dependencies} plugin builds its own representation of this Markov chain's probability matrix for every dependency, and then calculates the average percentage of time spent in each state (through the respective stationary distribution vector), which effectively represents the percentage of requests that the load-balancer will send to each replica based on the current conditions.
By combining this information with the quality score allocated to each replica/state in the previous step, the algorithm can accurately determine the average quality score experienced in a given node's communications with a dependency. This calculation is performed as the weighted average of the previously obtained scores, where the relative percentages of requests are used as weights.
Since some dependencies can have a larger impact on the performance of an application (or represent a more significant bottleneck to its operation) than others, developers also have the ability to specify a weight for each of the dependencies defined in the FogService resource. Using that information, the final node score is calculated as the weighted average of the various dependency scores. Algorithm \ref{alg:dependencies} summarises this process.

\begin{algorithm}
\scriptsize
\caption{Dependencies Scoring}\label{alg:dependencies}
\begin{algorithmic}
\State $nodeScores \gets \{\}$
\For{\texttt{each node in Nodes}}
	\State $nodeScore \gets 0$ 
	\State $latencies \gets \Call{GetNormalisedLatencies}{$Node$}$
	\If{$\Call{isEmpty}{$dependencies$}$}
		\State $nodeScore \gets 1$
	\Else
		\For{\texttt{each dep in dependencies}}
			\State $scores \gets []$
			\For{\texttt{each depPod in dep}}
				\State $podScore \gets (latencies[depPod] * depLatencyWeight) + (depMetrics[depPod] * depMetricsWeight)$
				\State $scores \gets \Call{append}{scores, podScore}$
			\EndFor
			\State $matrix \gets \Call{GetMarkovMatrix}{scores, dep}$
			\State $vec \gets \Call{GetDistributionVector}{matrix}$
			\State $weightedProb \gets \Call{matrixMult}{scores, vec}$
			\State $avgScore \gets \Call{matrixAdd}{weightedProb}$
			\State $nodeScore \gets nodeScore + (avgScore * depWeights[dep])$ 
		\EndFor
	\EndIf
	\State $nodeScores[node] \gets nodeScore$
\EndFor
\end{algorithmic}
\end{algorithm}

\section{Workload Runtime}
\label{sec:runtime}

\begin{figure}[ht]
	\centering
	\includegraphics[width=0.95\columnwidth]{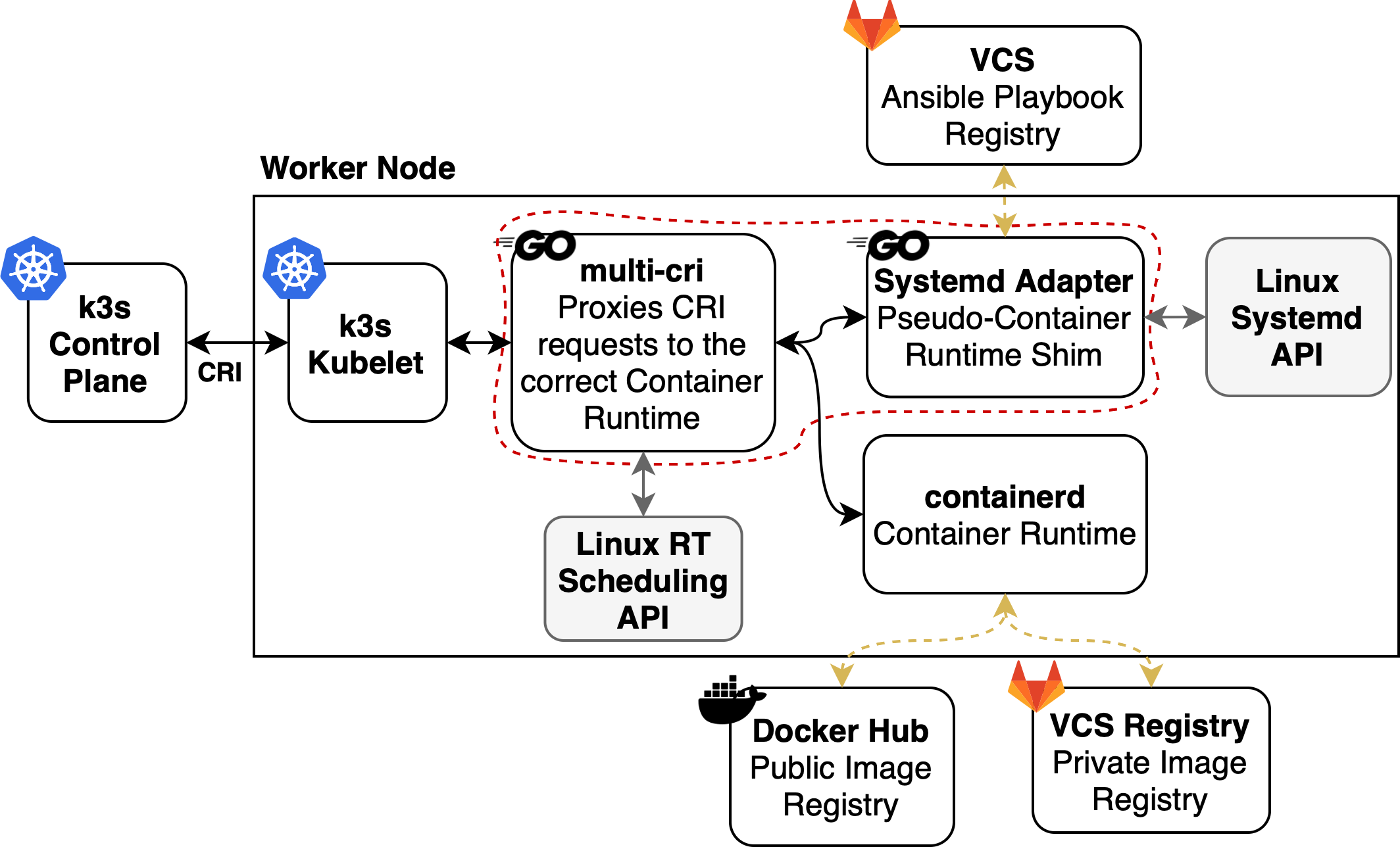}
	\caption{Workload Runtime Architecture}
	\label{fig:multi-cri}
    \vspace{-5mm}
\end{figure}

\subsection{Legacy Application Runtime}

Supporting the orchestration of legacy apps is an important requirement, since some services may not yet have been ported to run on containers, or may even be entirely incompatible with containerised environments due to, for instance, the lack of system access or excessive overhead.
After considering the available alternatives, we based our implementation on the Multi-CRI\footnote{\myref{https://github.com/atrioinc/multi-cri}} open-source Golang project, which acts as a proxy layer between the Kubelet and one or more runtimes.
To accomplish this, the Kubelet is configured to treat Multi-CRI as its container runtime, thus sending all \gls{cri} requests directly to it. The tool then selectively forwards these requests to the correct underlying runtime based on the $runtimeClass$ field associated with each Pod.

Our approach uses a custom version of this Multi-CRI tool and a purpose-built \gls{cri} Shim that converts the Kubelet's requests into $Systemd$ \gls{api} calls, thus leveraging $Systemd$'s complex process management architecture instead of re-engineering a similar solution. With this implementation, the Kubernetes control plane (as well as the developers and system administrators that use it) treats legacy applications exactly as regular Pods, while in the worker nodes themselves each Pod is, in reality, executed as a $Systemd$ unit/service. Any Pod without the 'legacy' $runtimeClass$ field is still treated as a normal containerised workload, as Multi-CRI forwards its \gls{cri} traffic to $containerd$ or whichever container runtime was configured for that particular worker node.

Notably, the current limitations of the system include the lack of network isolation and \gls{cni} integration. In the future, these features could be introduced through the use of Linux network namespaces.

\subsection{Attribution of Realtime CPU Scheduling policies}

In order to avoid the creation of another daemon service, and to leverage the fact that Multi-CRI already executes some logic each time a new Pod is deployed, the logic that enables automatic attribution and management of real-time \gls{cpu} scheduling policies is integrated within the customised version of the Multi-CRI used by the system.

To assign the priority value to each process, the real-time priority management module calls the \textit{chrt} Linux \gls{api} \cite{chrt_command} on the worker node itself. Within the FogService schema, real-time processes are specified either by their PID within the container, or by the process name (or a substring of it) that can be used to find it. As a consequence, the \textit{chrt} utility cannot be used directly, given that it requires the process' PID on the host namespace. 
To solve this problem, the module starts by using the \textit{nsenter} utility to enter the PID namespace of the container, while retaining the ability to use the host's \textit{chrt} \gls{api}. 
In cases where one or more of a Pod's processes have a delayed activation relative to the Pod's start time, the system will be unable to find them (and therefore set their priority) when the Pod is created. In these scenarios, the affected processes are registered to a pending queue which is revisited every 30 seconds until all the processes have successfully been allocated their priority.

Finally, the module is also capable of allocating SCHED\_DEADLINE real-time scheduling parameters to processes. That process is significantly simpler than the one that was just described, since there is no need to rank processes by their priority. The system merely applies the attributes as specified in the corresponding FogService resource.


A final design challenge to take into consideration is how to ensure that the \gls{cpu} usage limits specified by developers are also enforced for the processes that use Linux's Realtime scheduling features.
Some container runtimes (e.g., Docker Engine) have already introduced support for this feature by configuring the \textit{cpu.rt\_period\_us} and \textit{cpu.rt\_runtime\_us} $cgroup$ parameters, where the latter represents the maximum number of microseconds of \gls{cpu} time that can be allocated to the $cgroup$'s tasks during each period. 
However, this has not yet been included in Kubernetes' Pod specification or in the \gls{cri} request schema, meaning that, even if the underlying runtime supports these configurations, Kubernetes currently has no mechanism with which to instruct it to apply them.

To address this, the real-time priority allocation logic included in the customised Multi-CRI service uses the \textit{cgset} Linux \gls{api} to set the two parameters manually on the respective $cgroup$.
In terms of determining which values to set, the simplest approach would be to use the \gls{cpu} resource limit specified by the developer in the Pod (or FogService) resource, and convert it to the required units. For Pods that are entirely composed of real-time tasks, this would work exactly as intended. However, in cases where Pods contain a mix of both real-time and regular processes, this solution would effectively double the Pod's usage limit, since both types of workloads would have identical restrictions that would be accounted separately.
Therefore, for the purposes of this implementation, the regular \gls{cpu} limits apply exclusively to processes that use \gls{cfs} (as usual), and the FogService resource includes a separate field to specify real-time limits.

\section{Cluster State Monitoring}
\label{sec:cluster_state_monitor}

As discussed in section \ref{sec:architecture}, continuously monitoring the current cluster state is required in order to ensure that the allocations remain optimal, which may require the re-scheduling of active Pods into the most suitable node.
This problem becomes especially relevant within the context of this implementation, as scheduling decisions are influenced by, among others, communication latency and application performance, which constitute notoriously volatile metrics. After the initial deployment of a Pod, as time passes, the conditions that lead to the allocation of the corresponding node will inevitably change, potentially leaving the cluster in a sub-optimal, or even degraded, state for long periods of time.

To address these issues, our proposal includes the Cluster State Monitor component, which periodically submits each existing Pod to a simulated scheduling process in order to determine if it is still running on the most optimal node.
To achieve this, this component is based on the Descheduler\footnote{\myref{https://pkg.go.dev/sigs.k8s.io/descheduler}} project by Kubernetes' Scheduling Special Interest Group and also integrates a modified version of OpenShift’s \textit{capacity-analysis} open-source project\footnote{\href{https://docs.openshift.com/container-platform/3.11/admin\_guide/cluster\_capacity.html}{https://docs.openshift.com/container-platform/3.11/admin\_guide/ cluster\_capacity.html}}, this provides a local Scheduler implementation loaded with the custom scheduling plugins described in section \ref{sec:scheduler}, as well as a dummy version of the Kubernetes \gls{api}.

For each of the active Pods in the cluster, the algorithm starts by synchronizing the dummy \gls{api} with its real counterpart, thus building a local representation of the current cluster state. Crucially, the Pod under evaluation is excluded from this local representation, given that the objective is to ascertain the worker node where it would be scheduled to, if it was not already deployed in the cluster. 
The Cluster State Monitor then uses the dummy \gls{api} to simulate the re-scheduling of the Pod by running the scheduler algorithm in a dry-run configuration that does not apply any changes to the actual cluster.
By comparing the simulation's result with the current worker where the Pod is located, the plugin can accurately determine if there is a more optimal node, and, therefore, if the Pod should be evicted. Algorithm \ref{alg:better-node} summarises this process.

This approach is able to make eviction decisions that take into account complex scheduling constraints and domain-specific logic. This, in turn, ensures that the orchestration framework is able to detect and respond to changes in the performance of services and network links that occur naturally over time. This way, the system maintains optimal deployment placement and a balanced cluster state.
Additionally, this approach eliminates nearly all false positives by confirming that the scheduler will not simply reallocate the Pod to the same node. This is achieved without burdening the cluster's scheduler with a large number of additional operations, since a separate, local, copy is used.

\begin{algorithm}
\scriptsize
\caption{State Monitoring}\label{alg:better-node}
\begin{algorithmic}
	\For{\texttt{each node in nodes}}
	\State $pods \gets \Call{GetPodsOnNode}{$node$}$
	\For{\texttt{each pod in pods}}
		\State start $\gets \Call{GetPodStartTime}{pod}$
		\State result $\gets \Call{SimulateScheduling}{pod}$
		\State now $\gets \Call{GetCurrentTime}{$$}$
		\If{$result \neq node$ AND $now-start > THRESH1$}
		\State lastBackoffTime $\gets$ backoff[pod]
		\If{now $-$ lastBackoffTime$ > THRESH2$}
			\State \Call{EvictPod}{pod}
			\State backoff[pod] $\gets now$
		\EndIf
	\EndIf
	\EndFor
	\EndFor
\end{algorithmic}
\end{algorithm}

\section{Load Balancing}
\label{sec:load_balancer}

By default, Kubernetes' algorithm for balancing network requests between replicas of the same service strives for maximum fairness, meaning that, on average, each replica should receive the same amount of network requests as every other, $1/N$.
While this strategy helps ensure balanced resource usage between replicas, it will frequently result in sub-optimal performance in terms of overall request round-trip time. 
Since most edge computing clusters in smart-city contexts are spread out geographically, some replicas will invariably be much farther away from a given client application, in network latency terms, than others.
In these cases, using a fair algorithm like the one employed by Kube-Proxy means that the farthest Pods will receive the same amount of requests as the closest ones, which will dramatically increase the maximum \gls{rtt} values and negatively impact the system's ability to comply with performance guarantees.
Furthermore, in real production deployments, replicas will frequently exhibit uneven performance and resource headroom due to being allocated on nodes that are more resource-constrained or that are host to more interference-prone processes. As such, an ideal system should leverage metrics exposed by the worker nodes and the applications themselves, both in real-time or in historical trend formats, to ascertain the replicas that are most likely to provide the fastest response time.

Using the default Kube-Proxy behaviour as a starting point, we propose a modified approach where the aforementioned integrations can be accomplished by dynamically changing the probability values of the rules in each service-specific $iptables$ chain. 
The FogService operator continuously allocates a normalised score to each replica. 
The scores are then stored within the Status section of the respective FogService resource. As such, the load-balancer services that require this information can access the summarised version by performing a single Kubernetes \gls{api} request, greatly reducing congestion.

This customized Load-Balancer service starts each cycle by iterating over the list of Service resources declared in the cluster and obtaining the relevant latency and metric information. Each replica $i$ of the respective FogService is then allocated a score based on the following formula: 

\[ s_i = (mv_i * mw_i) + (lv_i * lw_i) \]

where $mv_i$ represents the normalised and aggregated metric score of the given Pod, and $lv_i$ represents the normalised latency score of the node where that Pod is located.
In turn, $mw_i$ and $lw_i$ represent the weight that was specified on the FogService resource for $mv_i$ and $lv_i$, respectively. This allows developers and administrators to fine-tune the balance between prioritizing network latency or processing performance, for each separate service.
Using these scores as weights, the load-balancer daemon can successfully determine the percentage of the network traffic that should be directed to each replica. However, the $iptables$ \gls{dnat} rule probabilities cannot be updated directly with these weight values since the kernel executes rules in sequential order, forming a Markov chain.
In this Markov chain, $P_i$ represents the probability of remaining in each state (not transitioning to the next state). This probability, presented below, depends on the probabilities of no transition in the previous states, and assumes the value 1 when $i = N$, since there is no next state to transition to. Therefore, the replicas must first be sorted by score in ascending order, and the following formula must be used in order to accurately calculate the probability of a given rule being selected when it is considered by the kernel:






\[
    P_i= 
\begin{cases}
	s_1,              & \text{if $i$=1}\\
    s_i * \frac{1}{\prod_{j=1}^{i-1} (1 - P_j)},& \text{if } 1 < i < N\\
   1,				& \text{if $i$=N}
\end{cases}
\]


3After this process is complete, the load-balancer updates the existing rules in reverse order, so that the highest scoring replica is the first to be considered. By default, this algorithm is executed once every 30 seconds, as a balance between guaranteeing responsiveness to changing cluster conditions, and avoiding excessive Kubernetes \gls{api} queries and network congestion.

One potential drawback of this implementation is its decentralised architecture where different worker nodes do not coordinate amongst themselves to avoid situations where certain replicas start receiving excessive amounts of traffic. However, when applications expose meaningful performance metrics, the system can automatically self-correct this behavior in real-time, by detecting the fact that certain replicas' metric values are worsening, and adjusting the Markov chain probabilities accordingly.
This methodology is inspired by related work done in \cite{phd_loadbalancer}, and expands upon it by including the application metrics as an additional decision factor, since the original work focused only on the latency of network links.

Within the scope of the broader orchestration framework presented in this work, this component is crucial since, beyond the reasons that have already been mentioned, the \textit{Dependencies} scheduler plugin and, by extension, the cluster state monitor, would be rendered almost entirely ineffective without it. In fact, scheduling applications as close as possible to replicas of their dependencies would not have the desired effect if worker nodes used the standard Kube-Proxy load-balancer implementation.

\section{Performance Evaluation}
\label{sec:performance_evaluation}
\subsection{Experimental Setup}

\begin{figure}[t!]
	\centering
	\includegraphics[width=\columnwidth]{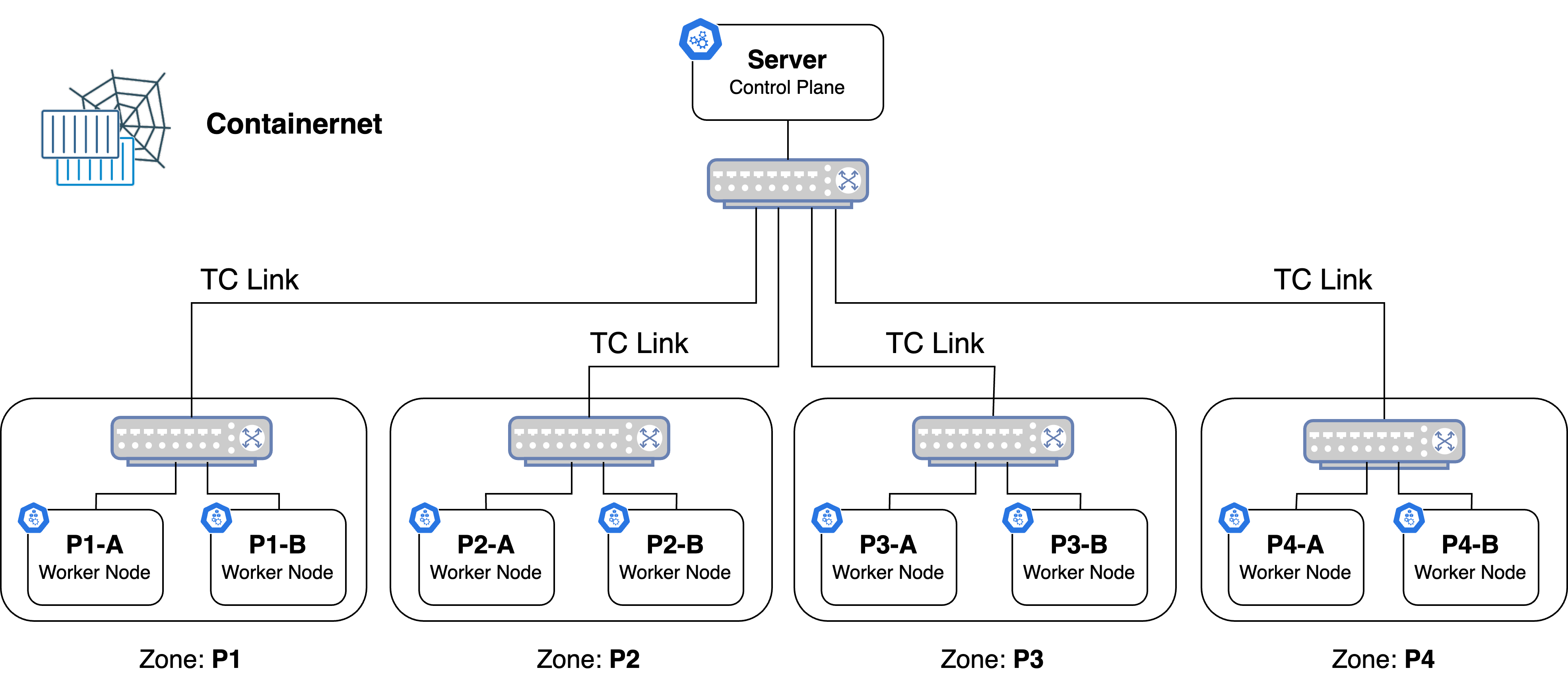}
	\caption{Overview of the virtualised cluster used in the evaluation setup}
	
	\label{fig:virtualization_architectures}
\vspace{-0.4cm}
\end{figure}

The functional and performance evaluation tests are conducted in a virtualised cluster framework presented in figure \ref{fig:virtualization_architectures}, which is deployed as a set of virtual machines running within a Proxmox Hypervisor. When compared to a platform using real devices instead of virtual machines, this approach enables deployments with a larger number of nodes and offers more flexibility regarding the cluster topology, networking architecture, and dynamic control of network link performance using \gls{tc}\footnote{\myref{https://tldp.org/HOWTO/Traffic-Control-HOWTO/intro.html}}. The cluster contains 1 master node and 8 worker nodes belonging to 4 different topology zones. All nodes $P4$ consist of Docker containers created from a customised Ubuntu image with $k3s$ and support for $systemd$.
The test environment is implemented inside $Containernet$ (based on Mininet)\footnote{\myref{https://containernet.github.io/}} to allow the creation of a dynamic number of containers, under a customised network topology. The performance of each link is configurable using \gls{tc}, which is useful to test scenarios where the links between the zone switches and the core switch have different latency values.

Furthermore, the evaluations explored in this section are mostly functional in nature, and aim to demonstrate and validate the behaviour of the implemented components in different scenarios. As such, the results mainly depend on the chosen cluster topology, network link performance, and characteristics inherent to the specific scenarios themselves, and not on the performance of the underlying worker nodes. \cite{artigo_NOMS} provides the necessary context regarding the performance of an Edge Computing SBC in several of the runtimes, priorities, and configurations used by this orchestration system.

\subsection{Scheduling Dependencies Plugin Analysis}

The goal of this test is to validate the Dependency plugin's capability to influence the Pod Scheduling process to prioritize the nodes, in order to provide the lowest amount of communication latency between the Pod and the best performing and/or most available replicas of its dependencies.
\begin{table}[]
\centering
\caption{Latency values configured in each network link}
\label{tab:dependencies_latency}
\begin{tabular}{@{}ll@{}}
\toprule
Network Link & TC configured latency \\ \midrule
P1 & 0.5 ms \\
P2 & 0.8 ms \\
P3 & 1 ms \\
P4 & 1.2 ms \\ \bottomrule
\end{tabular}
\end{table}

\begin{table}[]
\centering
\caption{Internal metric values for each destination node}
\label{tab:dependencies_metrics}
\begin{tabular}{@{}lll@{}}
\toprule
Dependency Replica & Static Metric Value & Worker Node \\ \midrule
dependency-0 & 5.0 & P1-A \\
dependency-1 & 1.0 & P2-A \\ \bottomrule
\end{tabular}
\end{table}
In this scenario, a dependency application has already been deployed to the cluster in the form of 2 replicas, dependency-0 and dependency-1. Table \ref{tab:dependencies_latency} indicates the latency values for the network links between each zone switch and the core switch. These latency values are induced on each link using containernet's \gls{tc} functionality. 
Additionally, the dependency application exposes an internal metric that the orchestrator can use to determine the current performance level of each replica. For the purposes of this experiment, the exposed values are statically defined and a lower value denotes a more suitable replica. These values can be consulted in table \ref{tab:dependencies_metrics}.
The experiment is conducted by a bash script that deploys the candidate Pod to the cluster, and then registers the node that was selected by the Scheduler, before removing the deployment to clear the cluster state. This procedure is repeated 200 times in order to ensure statistically relevant results. It is also important to note that the candidate Pod's image was pulled to each worker node in advance, so as to neutralize the effect of Kubernetes' \textit{ImageLocality} scheduling strategy, which favours nodes that already possess the required image.
Finally, each round of tests is performed in 3 different scenarios:
1) Using the default Kubernetes Scheduler without any indication of a dependency relation so as to establish a baseline behaviour;
2) Informing the default Scheduler of the dependency relation using the PodAffinity specification with Zones as the topologyKey; 
3) Using a custom Scheduler that includes the Dependencies plugin.

\begin{figure}[t]
	\centering
	\includegraphics[width=\columnwidth]{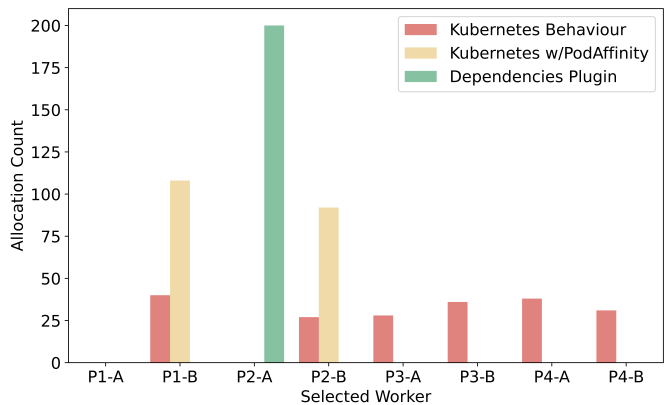}
	\caption{Distribution of scheduling results for each Scheduler configuration}
	\label{fig:c6_dependencies_results}
\end{figure}

Figure \ref{fig:c6_dependencies_results} presents the resulting distribution of the 200 scheduling attempts throughout the cluster's worker nodes, for each Scheduler configuration. The default Scheduler results demonstrate an overall trend for the equal (on average) distribution of allocations between the available nodes, which is consistent with the fact that the Scheduler does not possess any information concerning the dependency relation.
However, it is important to note the absence of any allocations to nodes \textit{P1-A} and \textit{P2-A}. This is attributed to Kubernetes' \textit{NodeResourcesBalancedAllocation} scheduling strategy, which disfavours those nodes due to the fact that they each already host one Pod (the  replicas of the dependency service), whereas the remaining nodes do not.
Compared to this first approach, the inclusion of the PodAffinity specification is a significant improvement, since the Scheduler becomes aware of the dependency constraint and applies a preference for nodes that are located within the same topology zone as a node with a running replica of the dependency. Nevertheless, results show that nodes \textit{P1-A} and \textit{P2-A} are still disregarded due to the same reasons as before, despite being the ideal choices. Furthermore, the plugin remains unable to distinguish potential nodes through the application-level metrics of the closest dependency replicas, trending instead towards an equal distribution of allocations between the nodes belonging to suitable topology zones.
The dependencies plugin resolves all of these issues by correctly identifying the \textit{P2-A} node as the optimal choice for the allocation of the candidate Pod, and consistently selecting it on all of the scheduling attempts, thus maximising the quality of the link between the candidate Pod and the service on which it depends.

\subsection{Scheduling Realtime Plugin Analysis}
\label{sec:realtime_results}

\begin{figure*}[t!]
	\centering
	\includegraphics[width=\textwidth]{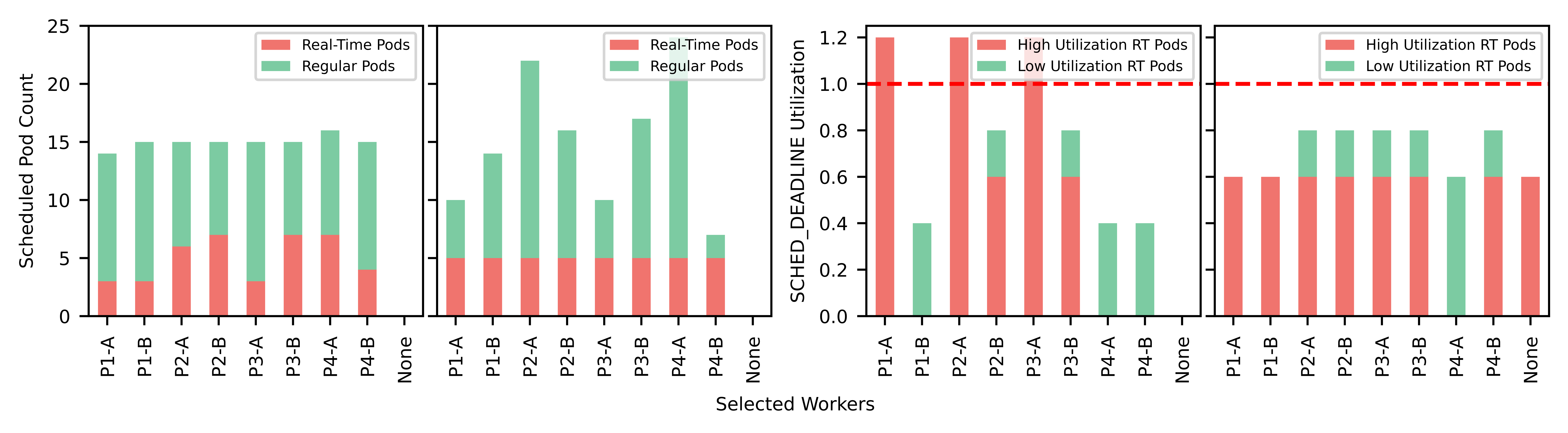}\caption{Number of regular pods and real-time pods allocated to each of the eight nodes for test suite number 1: default and proposed approach; and Total share of CPU time used by the Linux SCHED\_DEADLINE EDF scheduler in each node for test suite number 2: default and proposed approach}
	\label{fig:c6_realtime_results_1}
\end{figure*}

The goal of the test is to evaluate how both scheduler implementations take into account the real-time attributes of pods, and how well the final allocation balances these pods throughout the cluster.
In this scenario, 120 pods are declared at the same time. Of those, 80 consist of regular pods, and the remaining 40 pods possess real-time priority attributes; their deployment is restricted to the 8 nodes belonging to zones P1, P2, P3, and P4. 
The experiments are managed autonomously by a bash script which interacts with the Kubernetes \gls{cli} to perform the deployments of the test pods, wait until every pod has been scheduled, record the resulting allocation in a \gls{csv} file, and delete the deployments in preparation for the next test run. This process loops until all 20 test runs have been performed (for each of the two tests).   
The first set of graphs in figure \ref{fig:c6_realtime_results_1} presents the number of regular pods and real-time pods allocated to each of the eight nodes. The first subplot depicts the results achieved by using the default Kubernetes scheduler, whereas the data for the second subplot is obtained using the custom implementation. 
In order to facilitate the visual presentation of the data and the subsequent discussion, only one test run result is presented for each of the schedulers. An analysis of the remaining 38 test runs reveals that, while the precise values vary between runs, the trends and behaviours identified and discussed in this section are present in every one.

Results on the use of Kubernetes' default scheduler show that Kubernetes attempts to divide the load equally amongst the available nodes, with each receiving an average allocation of approximately 15 pods. 
Despite this, since the default scheduler does not possess the necessary context, it does not discriminate pods by their real-time priority or lack thereof. This leads to a pronounced imbalance in the number of real-time pods allocated to each of the nodes, which has implications for the performance of both types of pods, but especially for those with regular priority, as discussed in \cite{artigo_NOMS}.
Results in the second subplot demonstrate that a custom scheduler equipped with the Realtime plugin is capable of distinguishing between the types of pods and balancing real-time pods evenly throughout the cluster. In this case, the initial 40 \gls{rt} pods are distributed equally as 5 pods for each of the 8 worker nodes. 
This minimizes the amount of \gls{cpu} allocation that is devoted to \gls{rt} tasks on each node, which in turn increases the performance of both types of pods. However, the analysis of the results also shows that the number of regular pods  per node becomes severely imbalanced during this process, which is an undesired consequence that occurs due to a convergence of factors.

As presented in section \ref{sec:scheduler}, beyond the capabilities discussed thus far, the Realtime plugin also influences the scheduling of regular pods by favouring nodes with the lowest level of interference by \gls{rt} tasks, as measured by the number of \gls{rt} pods and their respective resource requests. 
Since the scheduling of pods is performed sequentially and the experiment queues pods in an arbitrary order irrespectively of their \gls{rt} status, there is a high probability that there are multiple moments during the scheduling process of the 120 pods at which there is at least one node that has a lower number of \gls{rt} pods than the rest, as not all \gls{rt} pods will have been scheduled yet.
At these moments, if one or more regular pods enter the scheduling process, the Realtime plugin will assign scores that favour the aforementioned node or nodes. Over time, this leads to the imbalance in regular pod placement observed in the final result.
Therefore, given enough entropy in the queuing of new pods, this undesired consequence will always be present, to a certain degree. 
In normal circumstances, this side-effect would call into question the usefulness of the Realtime plugin. However, as presented in section \ref{sec:cluster_state_monitor}, this proposal's architecture includes a Cluster State Monitor that will gradually solve this issue, as will be demonstrated in section \ref{sec:cluster_state_results}.


The second test's objective is focused on the allocation of pods with SCHED\_DEADLINE attributes. Both schedulers are evaluated for how many nodes, if any, are allocated unfeasible \gls{rt} task sets, and also the overall balance of SCHED\_DEADLINE utilisation between worker nodes. In this scenario, 8 high-utilisation \gls{rt} pods and 8 low-utilisation \gls{rt} pods are deployed. High-utilisation is defined as a pod that requires 60\% of the total capacity available for \gls{rt} tasks, whereas low-utilisation pods require 20\%. As such, if 2 high-utilisation pods are allocated to the same node, their combined utilisation will total 120\% of the nodes capacity, thus constituting an unfeasible task set. As before, the deployments are restricted to the 8 nodes belonging to zones P1, P2, P3, and P4.
The second set of graphs in figure \ref{fig:c6_realtime_results_1} presents the total share of \gls{cpu} time used by the Linux SCHED\_DEADLINE \gls{edf} scheduler in each node. The first subplot depicts the results achieved by using the default Kubernetes scheduler, whereas the data for the second subplot is obtained using the custom implementation. As before, only one test run result is shown for each of the schedulers. An analysis of the remaining 38 test runs also revealed results consistent with the ones presented.

Results on the use of Kubernetes' default scheduler show that Kubernetes attempts to divide the load equally amongst the available nodes in terms of the number of pods, with each receiving an exact allocation of 2 pods, regardless of their real-time scheduling attributes. As a result, 2 high utilisation \gls{rt} pods are assigned to the same node at least once. Since their combined utilisation surpasses their node's maximum available quota, Linux will not grant the requested \gls{rt} attributes to the pod scheduled last, which will greatly impact its performance.
Results in the second subplot demonstrate that a custom scheduler equipped with the Realtime plugin can leverage its knowledge of the \gls{rt} attributes declared for each pod within their FogService CRs to calculate their \gls{edf} \gls{cpu} utilisation, and compare it against the maximum value for each prospective node, taking into account the pods that have already been scheduled to each one. All nodes present similar final utilisation values, to the extent that such is possible, due to the fact that the plugin also strives to minimize that value in every node, so as to increase headroom and improve performance for other types of pods. This validates the plugin's Sort extension point component.

Despite this, the analysis of the second subplot also reveals, as before, an unintended consequence caused by a similar set of factors.
Due to the aforementioned arbitrary queue order and the sequential nature of the scheduler, 3 low utilisation \gls{rt} pods were allocated for the \textit{P4-A} node before the final high utilisation \gls{rt} pod could be scheduled. Due to this sub-optimal configuration, every node in the cluster had an utilisation of at least 0.6, which made it impossible to schedule that final pod.
This does, however, demonstrate that the Filter extension point component of the plugin worked as intended, by filtering out any nodes where it would not be possible to schedule the pod. 
Ideally, this scenario should result in each node executing one high utilisation pod and one low utilisation pod, which would result in a full deployment without any pods being deemed unschedulable. Unlike the previous tests, this issue cannot be fixed using the Cluster State Monitor as currently implemented, since this situation would require a new monitor plugin with purpose-designed algorithms. 
However, it is possible to achieve the optimal allocation by using the preemption capabilities of the Realtime plugin. To accomplish this, high utilisation pods would need to be declared with a higher priorityClass in their respective FogService CRs. With that change, the scheduler would be able to preempt 2 of the existing low utilisation \gls{rt} pods on the \textit{P4-A node} in order to place the final high utilisation pod, after which the preempted pods would be re-scheduled to other nodes.

\subsection{Cluster State Monitoring Analysis}
\label{sec:cluster_state_results}

This section presents the tests and results of the cluster-based plugins to allocate the services to the best nodes, and to improve the performance of degraded nodes.
The goal of this test is to evaluate if the Cluster State Monitor is capable of gradually resolving the imbalance observed in section \ref{sec:realtime_results}, and how the number of pods allocated to each node evolves over that period of time.
This test starts with the allocation of the same 120 pods (80 regular pods and 40 real-time pods) described in the aforementioned test. Results are collected using a bash script that interacts with the Kubernetes \gls{cli} to obtain the current allocation of pods at a rate of 1Hz, and append that information to a \gls{csv} file.   

\begin{figure}[t]
	\centering
	\includegraphics[width=\columnwidth]{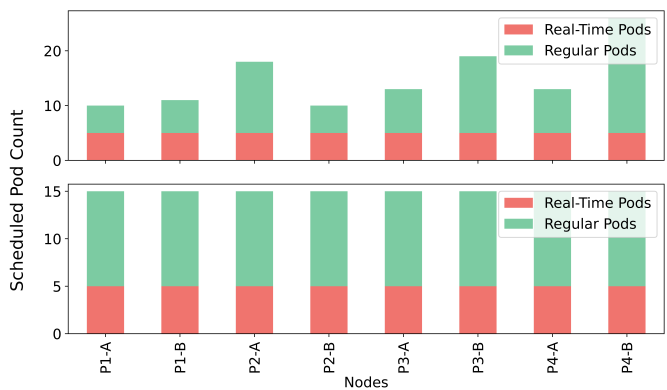}
	\caption{Number of scheduled  real-time pods and regular pods to each of the 8 nodes: initial state and after 380 seconds}
	\label{fig:c6_monitor_results_1}
\end{figure}

\begin{figure}[t]
	\centering
	\includegraphics[width=\columnwidth]{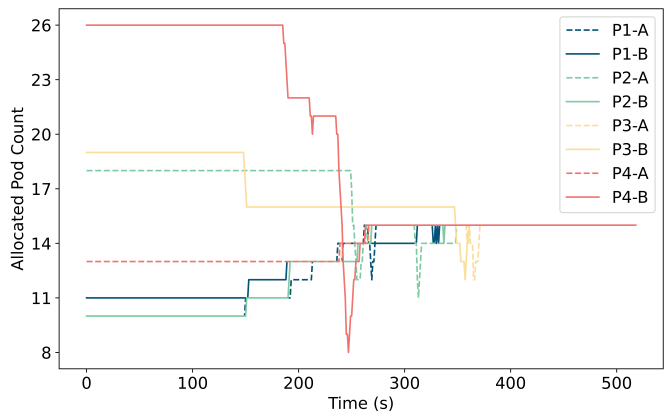}\caption{Evolution of the total number of pods (of both types) that were allocated to each node over time}
	\label{fig:c6_monitor_results_2}
\end{figure}

Figure \ref{fig:c6_monitor_results_1} presents the number of scheduled  real-time pods and regular pods to each of the 8 nodes. The first subplot depicts the initial cluster state, whereas the second subplot presents the cluster state after 380 seconds have elapsed. Figure \ref{fig:c6_monitor_results_2} presents the evolution of the total number of pods (of both types) that were allocated to each node over time.
Results in the second subplot of figure \ref{fig:c6_monitor_results_1} show that, during the 380 second period considered, the Cluster State Monitoring was able to completely re-balance the pod allocation of each node, in order to achieve an optimal state where both real-time and regular pods are perfectly balanced, with each node running exactly 5 and 10 pods of each type, respectively, totalling 15 per node. 
This result validates that this tool is able to successfully solve the issue presented by the Realtime scheduling plugin in section \ref{sec:realtime_plugin}.

By analysing figure \ref{fig:c6_monitor_results_2}, one can clearly discern the gradual nature of this process and the evolution of the result. 
Initially, no preemption events occur, since the entire set of pods was created at the same time, and the 120 second grace period is therefore still in effect. Once that time elapses, the cluster state monitor's descheduler loop determines that it must evict pods currently located in nodes that have been over-allocated, namely \textit{P2-A}, \textit{P3-B} and \textit{P4-B}.
Notice, however, that there are moments in which some nodes, like, for instance, \textit{P3-B} and \textit{P4-B}, present an excessive number of evictions compared to what was actually necessary, lowering the total number of pods on those nodes below the final value of 15. This behaviour occurs due to the fact that this tool operates by evaluating the current placement of each pod in the same sequential manner as the scheduler, as opposed to pre-computing the optimal sequence of evictions to obtain the ideal result in the least number of steps. 
At those instants in time, a number of evicted pods had not yet been re-scheduled, which lead the descheduler to attempt further balancing corrections that were ultimately not necessary. A pre-computed approach would yield more efficient results but at a greater computational cost, and likely requiring a very substantial increase in Kubernetes \gls{api} requests and Prometheus queries, which could present challenges in terms of scalability. Nonetheless, the Cluster State Monitor strives to minimize evictions as much as possible by using the backoff mechanisms described in section \ref{sec:cluster_state_monitor}.
Furthermore, these inefficiencies are most evident in scenarios similar to this one, where pods are added in bulk, which does not represent most real use cases, where sources of entropy that may require corrective action (such as the addition of new pods and cluster events like node reboots) mostly happen in small, gradual increments. 


\subsection{Request Load-Balancing Analysis}
\label{sec:loadbalancer_results}

The goal of this test is to evaluate how both load-balancing approaches (default and custom) take into account the network link latency between a client service and the server's various replicas, and the impact that those decisions have on the round-trip time of those network requests.
\begin{table}[]
\centering
\caption{Average RTT values for each destination node}
\label{tab:loadbalancer_latency}
\begin{tabular}{@{}lll@{}}
\toprule
Replica & Worker Node & Average RTT \\ \midrule
server-0 & P1-A & 0.043 ms \\
server-1 & P1-B & 0.063 ms \\
server-2 & P2-A & 2.763 ms \\
server-3 & P3-B & 3.165 ms \\
server-4 & P4-B & 3.578 ms \\ \bottomrule
\end{tabular}
\end{table}
In this scenario, a client application located in node \textit{P1-A} sends \gls{http} requests at a rate of 10Hz to a server which has 5 replicas spread throughout zones P1, P2, P3, and P4. Table \ref{tab:loadbalancer_latency} indicates the average link latency between the node hosting the client and the ones hosting each server replica, at the time of the test. These latency values are induced on each link using containernet's \gls{tc} functionality. 
Each server replica maintains a count of the number of requests it has received, and the client application records the round-trip time of every request in a \gls{csv} file.

\begin{figure}[t]
	\centering
	\includegraphics[width=\columnwidth]{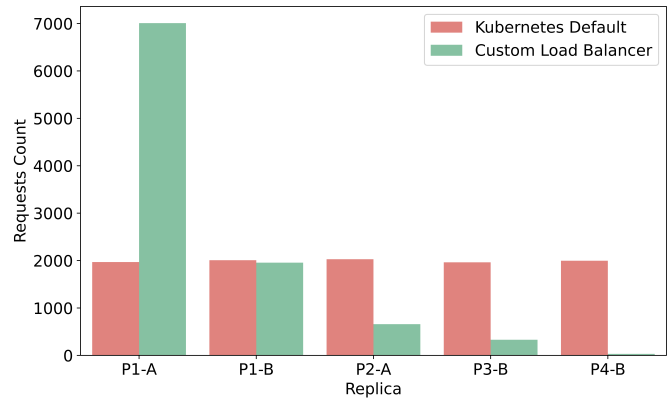}
	\caption{Number of network requests received by each of the five server replicas, in both tests}
	\label{fig:c6_loadbalancer_results_1}
\end{figure}

\begin{figure}[t]
	\centering
	\includegraphics[width=\columnwidth]{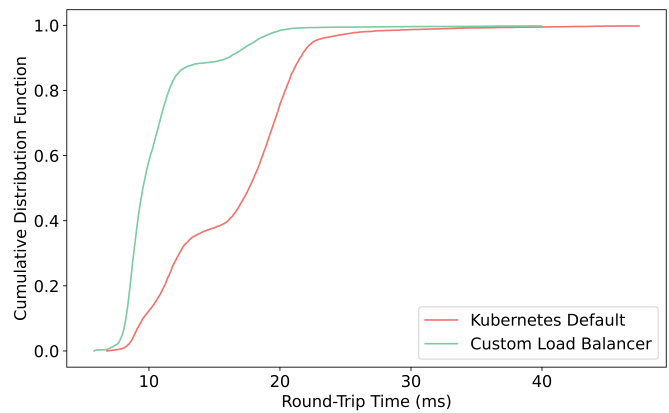}
	\caption{Cumulative distribution function of the round-trip time measurements taken by the client application, in both tests}
	\label{fig:c6_loadbalancer_results_2}
\end{figure}

Figure \ref{fig:c6_loadbalancer_results_1} presents the number of network requests received by each of the five server replicas, in both tests. Figure \ref{fig:c6_loadbalancer_results_2} presents the cumulative distribution function of the round-trip time measurements taken by the client application, also in both tests. Results on the use of Kubernetes' default load-balancing solution show that Kube-Proxy attempts to divide the load equally amongst the available replicas of the destination service in terms of the number of pods, with each receiving an average of approximately 2000 requests. 
Since the load-balancer does not possess the necessary context, it does not take into account the network link latency between the client and each of the replicas. As a result, both the average and standard deviation values for the round-trip time are high.

In contrast, results for the custom load-balancing solution demonstrate that it directs the vast majority of requests, 7000, towards the closest replica, which in this case resides within the same worker node. Other replicas receive increasingly fewer requests the farther they are located from the node executing the client application, in latency terms.
As a result, average and standard deviation values are much improved, as evidenced by the respective cumulative distribution function.
Note, however, that the custom solution still presents high maximum values, which correspond to the portion of requests answered by the farthest replica.

\section{Conclusions and Future Work}
\label{sec:conclusion}

This work comprised the development of a modular orchestration framework that extends
Kubernetes in order to improve on the current state-of-art in several key areas: 1) Abstracting the complexity of Kubernetes deployment specifications to encourage the participation of more stakeholders; 2) Influencing application placement based on performance and network metrics, \gls{rt} utilization, and the expected \gls{rtt} between dependencies, thus optimizing workload distribution; 3) Dynamically re-scheduling workloads in order to react to changes in application performance and link latency, thus ensuring stable performance; 4) Ensuring that critical workloads fulfill their time-constraints by automatically configuring RT scheduling policies; 5) Enabling Kubernetes to also manage legacy applications running natively, increasing the level of control and efficiency; 6) Improving the average RTT of network requests between services by performing load-balancing based on performance and network metrics. 
Overall, the components of this proposal accomplish the stated goals, facilitating a wide range of applications from, for example, critical C-ITS use cases for emergency teams where the fulfillment of time constraints is guaranteed, to 5G private networks with distributed User Plane Function (UPF) entities, where the client’s network traffic is forwarded to the instance of the destination service that is closest to its UPF.

The following aspects are not yet present or can be improved as future work:
1) Integrate the orchestration of the network layer and/or Time-Sensitive Networking technologies; 2) Implement support for live migrations using checkpointing techniques; 3) Integrate latency-sensitive and highly-available storage architectures; 4) Introduce mobility of the users and services in scheduling and load-balancing decisions.

\section*{Acknowledgment}

This work is supported by the European Regional Development Fund (FEDER), through the Regional Operational Programme of Centre (CENTRO 2020) of the Portugal 2020 framework and by National Public Funds through FCT I.P. (OE) [Project SNOB-5G with Nr. 045929 (CENTRO-01-0247-FEDER-045929)]

\ifCLASSOPTIONcaptionsoff
  \newpage
\fi



\bibliographystyle{IEEEtran}
\bibliography{bare_jrnl}
%

%








\end{document}